\documentclass[amsmath,amssymb,aps,pra,twocolumn,natbib]{revtex4}

\usepackage{graphicx}
\usepackage{dcolumn}
\usepackage{bm}

\begin{document}

\title{The Price of Gold: Curiosity?}

\author{Daniel W. Hook}
\affiliation{Digital Science, London, N1 9XW, UK}%
\affiliation{Centre for Complexity Science, Imperial College London, London, SW7 2AZ, UK}
\affiliation{Department of Physics, Washington University in St Louis, MO, USA}
\author{Mark Hahnel}%
\affiliation{Digital Science, London, N1 9XW, UK}%
\author{Christian Herzog}
\affiliation{Digital Science, London, N1 9XW, UK}%

\date{\today}

\begin{abstract}
Gold open access as characterised by the payment of an article processing charge (APC) has become one of the dominant models in open access publication.  This paper examines an extreme hypothetical case in which the APC model is the only model and the systematic issues that could develop in such a scenario.
\end{abstract}


\maketitle

\section{Introduction}
Imagine a world in which all scholarly articles are available through an open access route.  For many advocates of open research this is the dream: That every article produced by an academic at a publicly-funded research institution should be available not only to every other academic at every other research institution, but to the world at large. The philosophy that underlies this dream is a compelling one: that knowledge should be free, especially when paid for by the state and funded by the taxpayer.  This is not mere rhetoric, there are powerful arguments beyond the oft-mentioned financial ones that support this position.  If academic work is shared freely as a common good, then the chance of work being needlessly or unknowingly duplicated is minimized; access to and propagation of ideas may be faster; and, research impact may be greater. Importantly, the public who fund the research should not only have the right to benefit from the results of the research, but also have a right to have access to the outputs of research.

There are some fields where ``open access'' is considered to be a standard part of the research process, but the way in which open access is achieved remains diverse with a number of competing approaches. In parts of physics, mathematics and computer science arXiv (\url{http://arxiv.org}) is popular, in economics RePEc (\url{http://repec.org/}) is established, in bio science (especially genetics) BioRxiv (\url{https://www.biorxiv.org/}) has recently gained traction. Since the open access revolution in 2017 and 2018 a range of technology platforms have been used to launch a plethora of new subject repositories in diverse areas including chemistry, earth sciences, psychology, engineering, social science and beyond. While arXiv and RePEc were founded in the late 1990s and have had time to gain traction, they by no means have 100\% coverage in their respective fields.  The more recently launched subject repositories are likely to move faster in gaining acceptance due to the greatly increased visibility and awareness of open access.  However, there will still need to be time allowed for culture change to occur in each field before these mechanisms of sharing research are fully accepted. 

Platforms like these typically focus on a form of open access known as green open access (no processing charge is paid for the article to appear; the author typically chooses to make a  non-journal-branded version of a pre-submission, pre-accepted or pre-publication or even post-publication version of an article, depending on specific journal policies). Some journals, referred to as \emph{overlay journals} even use the arXiv infrastructure to run their journals, adding the missing refereeing functionality that arXiv does not provide. Sometimes these overlay journals are free to publish in, sometimes a payment is needed.  A rich ecosystem of services and publishing routes spanning green and gold open access and traditional subscription publishing exists around the arXiv and similar services.   Many of the services that are duplicating the arXiv model gain support from a mix of community funding, donations by philanthropic organisations, or from structured approaches that engage groups of academic institutions, and scholarly societies. The pure self-archiving activity on arXiv and similar platforms remains very much a grassroots movement in open access.

While these platforms have been transformative in demonstrating the potential of open access for the research community through sharing primary research content freely and instantly at low cost to the  community\footnote{At the time of writing the arXiv website (\protect\url{https://arxiv.org/help/support/faq##5A}) declared operating costs of \$1,465,000 per year and hosted 1,529,609 articles as of (2019-04-24).  Hence, a cost of just over \$1 per article per year.} it does not duplicate all the functionality of a formal research journal. While low cost options such as overlay journals do exist, they are still typically limited to specific research fields and it is still to be seen whether they are able to scale to meet broader demand (see for example, Quantum Journal, \url{https://quantum-journal.org/}). In general, arXiv and similar platforms to not offer a peer-review mechanism and as such lacks the ability to provide the prestige that academics need to progress in their careers in what is essentially a prestige economy \cite{Blackmore}. Offering peer-review (and through it eseteem), editing, selectivity/curation and an increasing range of services is where journals differentiate themselves from these platforms.

\begin{figure}
\centering
\includegraphics[width=0.45\textwidth]{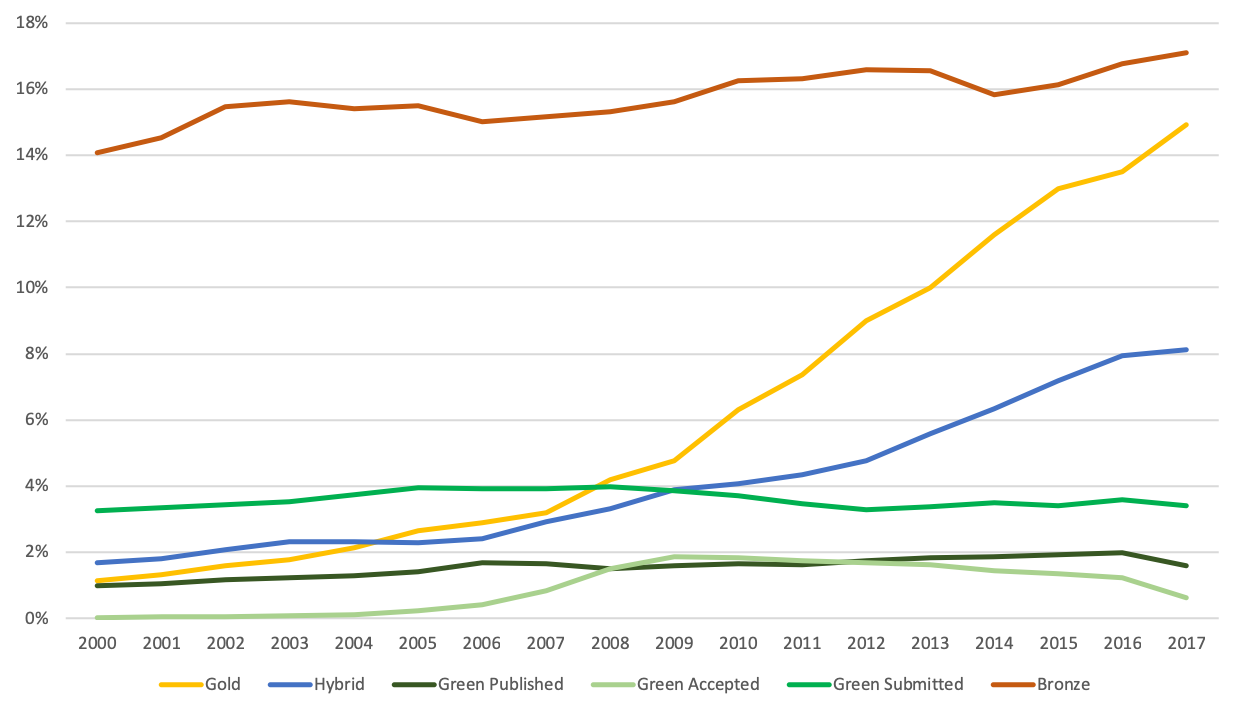}
\caption{\label{f1}Growth of open access content for articles in \emph{Dimensions} as a percentage of global research output between 2000-2017. While ``Bronze open access'' is the the prevalent route of free access, it is at the goodwill of the publisher and can be rescinded at any point. There is a steep growth in gold open access, a trend mirrored to a lesser extent in hybrid open access (likely to be due to the similarity of this mode with gold open access).  All forms of Green open access are essentially flat as a percentage of output. open access classifications described in Table \ref{t1}. Source: Digital Science \emph{Dimensions} \cite{HPH}.}
\end{figure}

\begin{table}[!h]
    \centering
    \begin{ruledtabular}
    \begin{tabular}{l p{5.5cm}}
       \textbf{Classification} & \textbf{Explanation} \\\hline
       Gold  & Version Of Record is free under an open licence from a full OA journal \\
       Hybrid  & Version Of Record is free under an open licence in a paid-access journal \\
       Green Published  & Free copy of published version in an OA repository \\
       Green Accepted  & Free copy of accepted version in an OA repository \\
       Green Submitted  & Free copy of submitted version, or where version is unknown, in an OA repository \\
       Bronze & Freely available on publisher page, but without an open licence\\
    \end{tabular}
    \end{ruledtabular}
    \caption{Classifications of open access shown in Figure \ref{f1}. Source: Digital Science \emph{Dimensions}.}
    \label{t1}
\end{table}

In contrast to the subject-based platforms mentioned above, geographically-aligned initiatives has tended to be less grass-roots and more top-down in nature. In a number of cases, these initiatives are coordinated with research evaluation infrastructures, as seen in the UK. The influential 2012 report by Janet Finch\cite{Finch} examined the UK's open access landscape and advocated continued publication in traditional journals but by paying the publisher a fee, allowing articles to be instantly freely available, the paid \emph{gold open access} model. This was recommended over mandating author-initiated deposit in either subject-based or institutionally-homed digital repositories and the then prevalent subscription-access model of academic publishing. Arguably Finch's recommendations established the \emph{Article Processing Charge} (APC) as the UK's preferred publication model. 

In essence, the report was trying to balance two competing needs of government: Firstly, the need increase the availability of research to the public which was paying for it was paying; and, secondly, the need to be able to evaluate research and researchers in a way that continued to allow public funders to make discriminating choices of who to fund, institutions to determine who to hire, and UK research to continue to be visible on a global stage.  In short, the UK system needed to make a choice that continued to align it with the existing prestige economy of journal-based peer review and could not cut ties with that world by mandating open access through a subject-repository route.  All of this stemmed from the aftermath of the financial crisis in 2008, which resulted in a subsequent need to increase accountability for public services in general.  This applied particularly to research funding as it had been protected on the basis that research made a positive contribution to the future economic growth of the UK, while other sectors faced significant cuts under the austerity policies that were introduced.  This accelerated the UKs move toward being a high-evaluation environment for research.

Since the turn of the millennium, and significantly since the publication of the Finch Report, the gold open access model, with the APC at its core, has risen sharply and has left Green open access lagging despite its greater cost (see Fig.\ref{f1}).

While managing a journal requires time and effort with even the leanest of open access journals needing to maintain a platform, source reviewers, manage a pipeline of content and ensure that the articles on the platform can be discovered and hence read by anyone who might be interested, profitability in a space that is principally publicly funded continues to be a contentious issue. On the one side there are those that believe that scholarly publishing is too lucrative an industry for publishers (for example, \cite{GM}) while on the other there are those who look to the future and who don't believe that there is enough money in publishing to handle all the publications that we might wish to publish \cite{green1}. It is important to consider the interaction of the models used to pay to support scholarly publication and communication with the prestige economy that drives academia.  This interaction is central to the way that academia currently functions and a major transition in the model that governs one part of the system (publishing), could produce unintended effects for the functioning of the other part (research evaluation, funding and career progression) that take many years to manifest and hence many years to remedy.

In this paper we conduct a \emph{Gedankenexperiment}: We determine the likely features of the research world in the case that the gold open access model became not just the prevalent but the only open access business model.  We add to our environment certain assumptions about the evaluation environment and citation behaviours. We then take the features that we discover through this experiment and compare with world as we experience it in 2019 to see if any of those features are exhibited in the current research environment.  We emphasize that although it is unlikely that the actual outcome of the open access movement should be an APC-only world, we believe that it is important to consider the features of such a world. The APC-model is fast becoming a dominant business model for publishers of not just the top journals but of journals throughout the ecosystem. 

We pass no judgement on the APC model itself but wish to understand the interaction of the APC model with institutional research strategies, research as a prestige economy as expressed through university rankings, promotion and tenure pathways and citation-driven evaluation.

The paper is organised as follows: In Sec.\ref{s1} we discuss the effects of the drive for research institutions to ``professionalize'' research over the last 30 years or so. In Sec.\ref{s2} we discuss the citation advantage of open access publications over closed-access publications and the interaction of this effect with academic success. In Sec.\ref{s3} we discuss the interaction of these effects, the potential outcomes of continuing to back a gold open access model as well as some other business models to support the sector's open access aspirations. Finally, in Sec.\ref{s4} we discuss possible approaches to circumvent the most extreme of the issues surfaced in previous sections.

\section{Strategy, Evaluation and Ranking\label{s1}}
\begin{figure}
\centering
\includegraphics[width=0.45\textwidth]{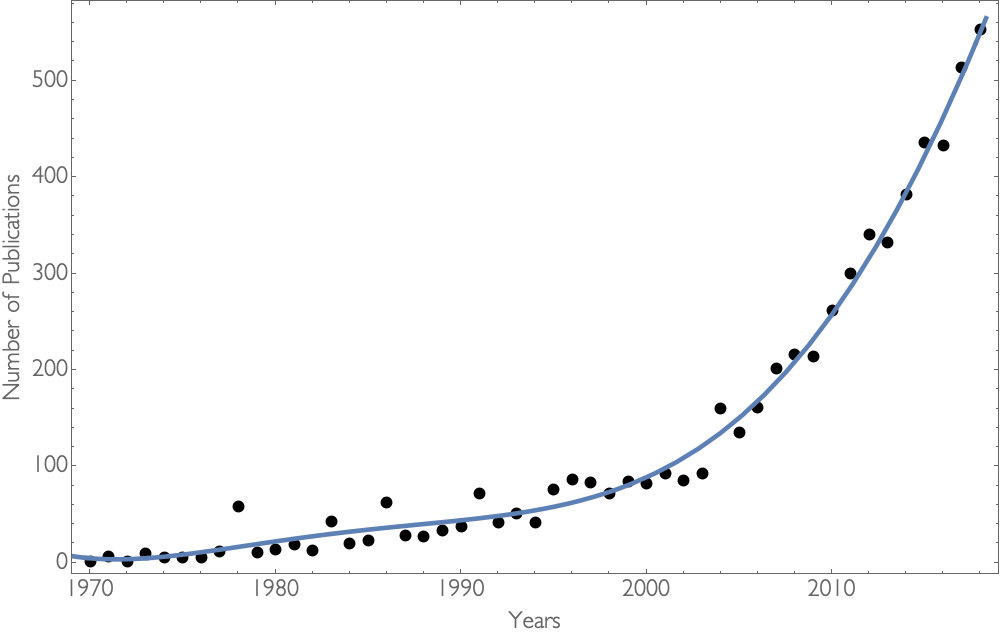}
\caption{\label{f2} The increase in articles mentioning the term "research strategy" in documents that also mention "bibliometrics", "scientometrics" or "research policy" as a full-text search, between 1970 and 2018. Black dots denote the raw data, the blue line is a curve of best fit to highlight the trend more clearly. Source: Digital Science \emph{Dimensions}.}
\end{figure}

The professionalisation of research has gradually progressed for more than 30 years.  Signs of this progression include the establishment of national evaluation exercises, university rankings, evaluation frameworks for funders, research strategy offices in institutions and the formalisation and public articulation of institutional research strategy.  We can quantify the increase in attention given to research strategy as a topic by examining the prevalence of the term research strategy in the relevant academic literature.  In Figure \ref{f2} we plot the increase in academic articles in the areas of bibliometrics, scientometrics and research policy that use the term ``research strategy'' since 1970.  We note a precipitous rise in the steepness of the curve that occurred since the mid-2000s that is well-correlated with the acceleration in professionalisation and evaluation frameworks in the post-financial-crisis era.  While the UK has been a leader in this space in its development of the REF 2014, it has been followed by many other countries \cite{Owens}.

The drivers for institutions to take on research strategies to generally adopt approaches that professionalise research are complex and not the focus of the current article.  However, we will note that even before governmental or funder-based drivers, institutions realised that while, in many countries, educational funding was linked to student numbers, research funding was linked to quality and volume.  Careful management of a research portfolio could bring in additional funding from many sources without the necessity to greatly expand infrastructure or operations, at least in the short term.

Internal pressures were quickly superceded in the UK by governmental interest in quantifying the quality and scale of research taking place in the country when, in 1986, Margaret Thatcher's government oversaw the an initial benchmarking exercise that gave rise to a second ``research selectivity exercise'' in 1989 and the establishment of the ``research assessment exercise'', the precursor of the REF, in 1992 \cite{THE}.  The UK is not alone in this with Australia quickly following suit in the late 1980s and many other countries now running some form of government reporting exercise for research \cite{NEUMANN2002721}.  

Even for those countries without government evaluation linked to core public funding, there has been, for more than 15 years the pressure of participation in international rankings \cite{SHJT}.  And, even prior to these international rankings were subject-based rankings and a variety of national approaches.  All of these serving to put pressure on universities to in part defend the public money that they receive.

Hence, two of the biggest topics for most institutional strategy and planning offices are: rankings and evaluation.  This translates into a significant amount of senior staff time in institutions is dedicating to managing external perceptions of the institution in order to ensure student numbers or satisfaction, research volume, quality and relevance. As a result, there is a fundamental alignment between those who evaluate research institutions and the teams running those institutions.  A high-evaluation environment naturally exhibits measurement-driven behaviours.  Thus there is a natural trickle-down effect from government policy to institutional leadership, departments, and research groups.  National strategies translate to institutional, departmental and research group strategies.

The number of research institutions in the UK that articulate either an institutional or departmental level is significant.  In Table \ref{t2} we have summarised the public positions available on university websites for the Russell Group of research intensive universities in the UK.  

It is clear that in a world of professionalised research, having a research agenda for an institution that can be articulated to the public at large seems like a sensible idea that it well aligned with government and taxpayer accountability.  However, it is not clear that a thematic approach to the articulation of a research agenda is sensible or indeed necessary to deliver on the level of accountability that might be expected either by government or the public.

Three Russell Group institutions have chosen not to share a thematic strategy on their websites (denoted "General" in Table \ref{t2}), but rather, have chosen to express their aims, ethics and the values behind their research choices.  They outline their approach to attracting the best research talent, the communities that they serve and the impact that they seek to make in broad terms.  They often choose to highlight examples of excellent research on their site but not talk about research aims in a thematic way.  Of course, we have only examined their public articulation of their strategy through their website, and there may in fact be a themed research approach which is not shared with the public but even this change in emphasis is interesting to note.

Most Russell Group institutions do express a set of strategic research themes either at an institutional or departmental level.  This may not be reflective of an institutional interpretation of a national strategy but through evaluation activities, competitive funding calls and a host of other small measurements, it is clear that successful institutions are typically seeking to ensure those who are evaluating them understand their research strengths in a subject-based manner.

We have evidence that measurements can be tremendously powerful in high-evaluation environments.  One of the best examples is, in fact, the progression of open access publication in the UK.  Centralised approaches from government in the form of advice from both the research councils (now forming UKRI), HEFCE in the form of the REF guidance and investments from Jisc have shaped the UK research environment. The uptake of open access for the UK has been a successful strategy for keeping the UK's research at the forefront of global attention as it continues to be a preferred collaboration partner and produces some of the world's highest cited content \cite{HHC}. Strategic alignment is a powerful tool to shape behaviour in high-evaluation environments.

\begin{table}[!h]
    \centering
    \begin{ruledtabular}
    \begin{tabular}{p{5cm} c}
       \textbf{University} & \textbf{Level} \\\hline
       Durham University   & Institutional \\
       Imperial College London   & Institutional \\
       King's College London   & Institutional \\
       London School of Economics and Political Science (LSE)  & Departmental \\
       Queen Mary University of London & General \\
       Queen's University Belfast & Institutional \\
       University College London   & Institutional \\
       University of Birmingham   &  General \\
       University of Bristol   & General \\
       University of Cambridge   & Institutional \\
       University of Cardiff   & Departmental \\
       University of Edinburgh   & Institutional \\
       University of Exeter   & Institutional \\
       University of Glasgow   & Institutional \\
       University of Leeds   & Institutional \\
       University of Liverpool   & Institutional \\
       University of Manchester   & Institutional \\
       University of Newcastle   & Institutional \\
       University of Nottingham   & Institutional \\
       University of Oxford   & Departmental \\
       University of Sheffield   & Departmental \\
       University of Southampton   & Institutional \\
       University of York   & Institutional \\
       Warwick University & Institutional\\
    \end{tabular}
    \end{ruledtabular}
    \caption{Classification of university public articulation of research strategy for Russell Group institutions in the UK.  Data obtained by searching university websites on 2019-04-24, locating research strategy pages and documentation and assessing whether research is organised into research themes, beacons, or a similar construct at a: a) Institution level; b) Departmental level; c) Not at all (General).  In the case that there was an articulation at a Central level, there may additionally have been Department level articulation of research themes, but if noted that there was a Departmental level, then the Central level was absent or could not be found.}
    \label{t2}
\end{table}

\section{The OA Citation Advantage\label{s2}}
In the professionalised research environment evaluation is commonplace either at an institutional, departmental or personal level.  Different geographies see different preferences: In the United States, national evaluation in the form of a REF or ERA style activity does not exist, but evaluation of individuals is commonplace and lies at the centre of both for the tenure and promotion process and also for obtaining research funding from national, local or charitable funding bodies.  In the UK, Australia, New Zealand, Japan and an increasing number of geographies, national evaluations form part of the research agenda.  Even in countries were formal evaluation is low, informal evaluation through university ranks continues to impose some pressure on how research is done.

\begin{figure*}
\centering
\includegraphics[width=0.9\textwidth]{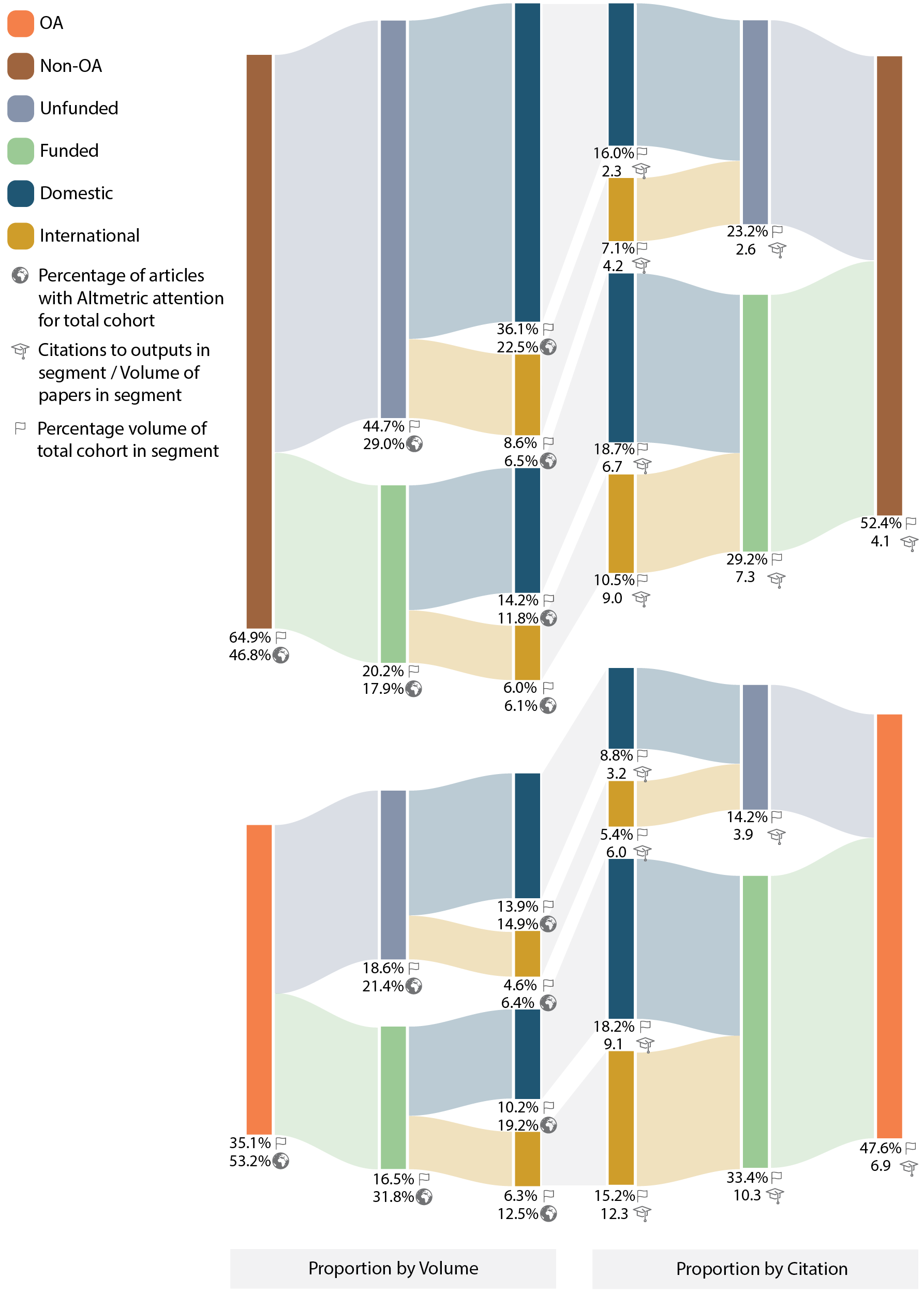}
\caption{\label{f3} Comparison of open access volume with open access citation rates.  On both sides of the diagram, the exterior categorisation shows the proportion of open access to non-OA content produced by the top 12 OA producing countries in 2016.  Stepping in one level, open access is split into funded and unfunded publications (acknowledgement made to a funder in the text of the paper).  Finally, the interior classification splits into internationally collaborative (more than one country affiliation in the list of co-authors on the paper) and domestically collaborative (only a single country represented). Reproduced by permission of Digital Science from \cite{HHC}. Source: Digital Science \emph{Dimensions}.}
\end{figure*}

The driver for all of these exercises is data.  The quantities that can be measured in a fair way across the research spectrum are limited: number of grants won, amount of funding won, number of paper published and number of citations to papers associated with a specific institution are the main measures available today with sufficiently low cost that they are feasible to be used in an evaluative setting.  Journal Impact Factors have been used as a proxy for the quality of the research contained in a paper for some years. Many have argued the inappropriateness of this practice (see for example, \cite{adams} and references therein).  

In more recent years article-level metrics have become more available and are becoming a more popular mode of assessing a paper.  There is also an increasing realisation that these metrics do not assess quality so much as attention associated with a paper.  Citations may be made for many reasons: For background or history of an idea; indicating that prior work is foundational in the current paper; highlighting an established result; questioning an established result or even showing that an established result is incorrect or has now be superceded. Mixing all these different rationales together and badging this as a mark of quality does not make sense.  And, while this is understood by academics, administrators and evaluators in general, citation counts or some normalisation thereof are usually contained within the basket of metrics provided to a peer reviewer in most evaluation processes.  Success in funding is often either subliminally or explicitly linked to citation performance.

Advocates of open access list the citation advantage as one reason to engage with this citation mechanism.  Indeed, the statistics for UK-based research benefit significant from the enhanced levels of citation garnered by open access articles.  In a recent report, an attempt was made to quantify the citation advantage achieved through various different drivers (see Fig.\ref{f3}) \cite{HHC}.

Figure \ref{f3} shows that just 35.1\% of the publication volume of the top 12 open access producing countries results in 47.6\% of the citations to papers in 2016, with the average open access publication getting 6.9 citations between its publication and the end of 2018, when the data were compiled.  This compared with 64.9\% of publications that were closed access receiving just 52.4\% of the citations, translating to the average closed access paper receiving 4.1 citations in the same period.  That's a 68\% advantage for citation success, just for being made openly available.  If other factors are taken into account such as whether a paper is funded and whether it is internationally collaborative, then the lead stretches further:  A funded, internationally collaborative, open access paper published in 2016 will on average, at the end of 2018 have received 12.3 citations versus just 2.3 citations for its closed access, unfunded, domestically collaborative counterpart, on average.  That's a more than 5-fold advantage.

\section{Thought experiment\label{s3}}
Given the background elements described in prior sections we can now present our thought experiment. Let us imagine that we are postdoctoral students who have just newly joined institutions in a world were the gold open access APC model is not only the prevalent mode of publication but the only mode of publication.  Additionally, we will assume a high-evaluation, professionalised research environment in which our host institutions have published thematic research goals on their websites.  Further, let us assume that our funders did not make provision for publication charges in our grants and that we are dependent on a centralised university fund for our ability to publish in ``top journals''.  In what position do we find ourselves?

Most research institutions will hire postdocs based on ability and track record during their PhD.  However, this is still a very early stage in a research career and so there is a significant risk associated with with hiring at this stage.  In many subjects, postdocs do not have their own research agenda and are rather joining a team to work on some larger project.  However, there are still a significant number of fields where the postdoctoral stage of a research career is where you have your first freedom to research an area and to let your curiosity roam.  It is the time during which future principal investigators are able to grow up and focus on their own projects.

Yet, in the world painted here, what degrees of freedom are available?  Without funding of our own, and relying on the institution's central fund for open access publication it is true that many institutions will be mindful of those in their early career and seek to ensure that we have the freedom to publish at least a few papers that are in line with our research interests.  But, many institutions themselves do not have that much freedom.  Funding is sufficiently tight that they must be pick carefully what works are published in open access journals.  They must choose not only which articles will make the biggest difference for them but also from a strategic perspective, which articles will contribute to the best evaluation results, and which serve to contribute to a coherent picture of the institution.  This means that if we are to publish then it must be in an area that is aligned with the institution's research themes.  If it was clear before joining the institution that this would not be the case then it may even be that a job would not have been made in the first place.  Hence, the cost of the APC model in this extreme version of a research environment is a significant reduction in our capacity to carry out curiosity-driven research.

The practical reality of the situation here is clear. However, the impact on the academic freedom of curiosity-driven research by the interaction between a high-evaluation environment and an APC-dominated publication landscape is not intuitively obvious at first glance. The loss of this particular academic freedom at this early point in a career, before an academic is established with their own grant, is of significance to the sector as a whole.  Much curiosity-driven research is only carried out during the postdoctoral period before academics gain broader responsibilities such as maintaining grant incomes, developing and managing their research group or  increased teaching and administrative workloads.

A further iteration on our thought experiment could be to consider what might happen if the role of peer review in research evaluation is decreased to reduce the costs of research evaluation in favour of a more metrically driven approach. In this case, we will step back from an exclusively APC scholarly communication model and consider a world in which the APC is a merely a significant or dominant publication model.

In this case, we might assume that citations rather than impact factors would become the currency of evaluating attention, which might erroneously be used as a proxy for quality.  In this scenario, articles published through an APC route would gain, on average, more citations than their closed-access competitors as shown in Sec.\ref{s3}. This means that subject areas with the money to publish through open access routes would have a route to gain proportionally more citations than those areas without such funding.  Intrinsically, this sets up a vicious circle that marginalises subject areas that have lower funding levels. Hence, even within disciplines that are well-funded, popular rather than good areas come to the fore at the cost of areas where open access is less available.  Hence, the cost of the APC model in this context is diversity of research disciplines.  The areas most likely to be effected are initially Arts, Humanities and Social Sciences, simply based on relative funding levels.  But, over and extended period, we hypothesize that it would be possible to see deficits in diversity across many fields.

\section{Discussion\label{s4}}

The thought experiment above shows some possible futures for the direction of the research ecosystem, albeit in extreme situations.  Nevertheless, the authors are sure that some readers will have seen situations that are close to some of the scenarios considered above in the world as it is now.  One, perhaps trivial, observation is that mono-cultural systems such as the one described here in the form of a single paid gold open access model for scholarly communication does not sit well with academic freedoms.  Indeed, we have seen the effects of the monopoly of single data sources and single metrics on the research system first hand in other contexts.  A diverse set of options that take account of different behaviours within different fields, and different funding models associated with different disciplines would seem to be healthy.  It is perhaps this observation that is, in part, behind the proliferation of different technologies that support the different subject-based green open access repository platforms.

In one scenario above the cost of an APC-exclusive world was the freedom to do curiosity-driven research, unless institutions were run a sufficiently enlightened manner or they had sufficient funds to grant that freedom.  In a real-world scenario, one could imagine that this might lead to a highly-inhomogeneous environment depending greatly on the nature of the management of the institution that you happen to do research at.  One can further imagine something of a free-market model arising in this world where institutions become highly competitive rather than collaborative.  This world also potentially favours the rise of large companies such as Google, Apple or Tencent as significant players in the research space since they would have both the money and the long-term view to allow researchers to undertake curiosity-driven research.  Of course, these companies are under no mandate to make the research that they fund openly available and hence much of the supposed advantages of open access could be lost in such a world.

The other scenario that we discussed is perhaps easier to handle.  The of the scenario was the overall diversity of research. In this world, funders need to be incentivised to seek diversity of research so that the metrics that they track are sufficiently complex and varied that they don't cause this narrowing effect.  While there is an intrinsic problem  with the nature of the measurement scheme, the mono-culture problem is not as extreme in this case and hence can be tackled more easily.

In terms of the world that we actually live in, there are several projects that start to move research beyond the limitations of current set of scholarly communication models.  Projekt Deal-like approaches seek to ensure homogeneity of access to publishing in a way that circumvents some of the downsides of the APC model highlighted above, but in doing so causes different problems for publishers regarding the sustainability of their own business models when considering fairness of the cost distribution at national levels.  A number of philanthropic-research interested foundations and government funders have explored publication models available from participants such as Faculty of 1000 (F1000), where the funder pays directly for publication of researchers to whom they have given grants; or Figshare where institutions or funders pay for a platform for research to be made widely available.  The Austrian Science Fund (FWF), the Max Planck Society, CERN and a host of other institutions have supported SciPost's (\protect\url{http://scipost.org}) development of a minimal-cost peer reviewed platform\footnote{Scipost supporters: \protect\url{http://scipost.org/sponsors}} \cite{scipost}.

In the case of the F1000 model, publication is instant and peer review takes place after publication.  In models such as that of figshare or arXiv, either the community or the institution is responsible for the quality of material that is associated with their brand.  However, the synergy between these approaches suggests an intriguing approach to future research publication.

Given that the nature of the research article is changing and moving more toward the type of paradigm that we are used to with software, it may make sense to have a shared infrastructure that is owned collectively by the community for the publication of research outcomes.  Costs for the maintenance of the platform could be shared fairly across institutions, governments and funders.  This is not a mono-cultural system but rather the establishment of a new communication standard and would be the analogue of the structure that we currently take for granted with the research publications of today (that an article has a title, co-authors, institutional affiliations, a date of publication etc).  The reason that it is not a mono-cultural device is that diversity could exist on top of the standard.

In this case, we have to determine from scratch what the analogue of a research article needs to look like and how the esteem or prestige system works on the new platform. It may be that the brand of the institutions associated with a piece of research would be a proxy for review as a piece of work would first be published.  However, one could imagine a vibrant ecosystem of approaches that help to establish quality of research on this platform, each with a new business model attached.  For example, instead of the peer review system that we have today, the next evolution of peer review may include some level of AI/automated checks on particular types of research output to give a basic level of confidence in results.  More advanced checking could be provided by human peer reviewers at a cost and in return for a peer review badge with underlying notes.  Publishers in this world may be the purveyors of highly-recognisable or highly-esteemed badges since they would have established routes to reviewers and would need to develop efficient editorial filters for the vast amount of content residing on the platform.

While there are challenges associated with the changes that we are seeing in scholarly communication today, there is, at least, significant diversity in the environment.  If the thought experiment in this article is a helpful device in anyway then hopefully it shows that maintaining certain levels of diversity at certain points in the ecosystem is critical.\\

\section*{Author contributions}
All authors contributed equally to this work.

\section*{Funding}
Funding for this work was provided by Digital Science.

\section*{Conflicts of Interest}
DWH and CH are employees of Digital Science; MH is an employee and shareholder of Figshare. Digital Science is a shareholder in Figshare. Digital Science's 100\% owner, Holtzbrinck Publishing Group, owns a 53\% share in the publisher Springer Nature.

\bibliography{bibfile}

\end{document}